\documentclass[12pt,authoryear]{elsarticle}
\journal{Social Networks}
\pdfoutput=1

\usepackage[top=2cm,bottom=2.65cm,left=2.65cm,right=2.55cm]{geometry}

\usepackage{textpos}
\usepackage{algorithm}
\usepackage{algpseudocode}
\usepackage{listings}
\usepackage{color}
\usepackage{textcomp}
\usepackage{pslatex}  
\usepackage{graphicx,xcolor}
\usepackage{epsfig}
\usepackage{epstopdf}
\usepackage{pgf}           
\usepackage{amsmath}
\usepackage{amssymb,amsfonts,amsthm,breqn}
\usepackage{longtable}
\usepackage{pslatex}      
\usepackage{verbatim}
\usepackage{float}
\usepackage{appendix}
\usepackage{array}
\usepackage{varwidth}
\usepackage{multirow}
\usepackage{tabularx}
\usepackage{framed}
\usepackage{lipsum}
\usepackage[round]{natbib}
\usepackage{indentfirst}
\usepackage{longtable}
\usepackage[utf8]{inputenc} 
\usepackage{booktabs}
\usepackage{lscape}  
\usepackage{rotating} 
\usepackage{graphicx,fancyvrb}
\usepackage{setspace}                      
\usepackage[hidelinks]{hyperref}
\usepackage[T1]{fontenc}
\usepackage{indentfirst}
\usepackage{lineno}
\usepackage{etoolbox}
\usepackage{mathtools}
\usepackage{fixltx2e}
\usepackage{bigints}
\usepackage[normalem]{ulem}
\usepackage{booktabs}

\newcommand*{\AdjustMargins}{%
    \setlength{\beamer@rightmargin}{0em}%
    \setlength{\beamer@leftmargin}{0em}%
}

\newcommand{\thickhline}{%
        \noalign {\ifnum 0=`}\fi \hrule height 1pt
        \futurelet \reserved@a \@xhline
}	
		
\newcolumntype{"}{@{\vrule width 1pt}}

\def\infinity{\rotatebox{90}{8}}

\def\Hcal{\mathcal{H}\,}
\def\htheta{\hat{\theta}}
\def\thetas{\theta^{\ast}}
\def\rset{\mathbb{R}}

\graphicspath{{Figures/}}	

\allowdisplaybreaks

\journal{}
\begin{document}

%====================================================================================
%====================================================================================
\begin{frontmatter}

\title{Efficient Bayesian inference for exponential random graph models\\ by correcting the pseudo-posterior distribution}

\author{Lampros Bouranis\corref{corresp}}
\ead{lampros.bouranis@insight-centre.org}
\cortext[corresp]{Corresponding author. Address: Insight, O'Brien Centre for Science, Science Centre East, University College Dublin, Belfield, Dublin 4, Ireland}

\author{Nial Friel}
\ead{nial.friel@ucd.ie}

\author{Florian Maire}
\ead{florian.maire@ucd.ie}

\address{School of Mathematics and Statistics \& Insight Centre for Data Analytics, \\ University College Dublin, Ireland}

%====================================================================================
%====================================================================================
\begin{abstract}
Exponential random graph models are an important tool in the statistical analysis of data. However, Bayesian parameter estimation for these models is extremely challenging, since evaluation of the posterior distribution typically involves the calculation of an intractable normalizing constant. This barrier motivates the consideration of tractable approximations to the likelihood function, such as the pseudolikelihood function, which offers an approach to constructing such an approximation. 
Naive implementation of what we term a pseudo-posterior resulting from replacing the likelihood function in the posterior distribution by the pseudolikelihood is likely to give misleading inferences. We provide practical guidelines to correct a sample from such a pseudo-posterior distribution so that it is approximately distributed from the target posterior distribution and discuss the computational and statistical efficiency that result from this approach. We illustrate our methodology through the analysis of real-world graphs. Comparisons against the approximate exchange algorithm of \cite{caimo} are provided, followed by concluding remarks.
\end{abstract}

%====================================================================================
%====================================================================================
\end{frontmatter}

%====================================================================================
%====================================================================================
\section{Introduction}\label{section:intro}
The study of networks is central to a broad range of applications including epidemiology (dynamics of disease spread), genetics (protein interactions), telecommunications (worldwide web connectivity, phone calls) and social science (Facebook, Twitter, LinkedIn), among others. The high--dimensionality and complexity of such structures poses a real challenge to modern statistical computing methods.

Exponential random graph (ERG) models play an important role in network analysis since they allow for complex correlation patterns between the nodes of the network. However this model presents several difficulties in practice, mainly due to the fact that likelihood function can only be specified up to a parameter dependent normalizing constant. This impacts upon maximum likelihood estimation  which is difficult to perform for larger networks, where the full likelihood function is available but it is just too complex to be evaluated. \cite{robins2} presented various approaches for overcoming this problem.

The Bayesian paradigm has been used to infer Exponential random graph models. The challenge of carrying out Bayesian estimation for these models has received attention from the statistical community in the recent past \citep{caimo}. The main challenge encountered by such a Bayesian setting is the evaluation of the posterior that typically involves the calculation of an intractable normalizing constant. Sampling from distributions with intractable normalization can be done via Markov chain Monte Carlo methods (MCMC) and especially with the celebrated Metropolis--Hastings algorithm \citep{metropolis1953equation,hastings}.

Nevertheless, the normalizing term in the ERG probability distribution is a function of the model parameters and thus does not cancel as usual in the standard Metropolis--Hastings acceptance probability. This gives rise to a target distribution $\pi(\theta\mid y)$ which is termed \textit{doubly--intractable} \citep{murray}, where at each iteration of the Markov chain Monte Carlo scheme intractability of the normalizing term of the likelihood model within the posterior and intractability of the posterior normalizing term must be handled. More sophisticated Markov chains, such as the Exchange algorithm \citep{moller,murray} have been proposed to sample from those doubly--intractable targets. Here again, these methods are not directly applicable in the ERG context as they require independent and identically distributed (iid) draws from the likelihood, which is not feasible for this type of models.

This motivated the Approximate Exchange algorithm of \citet{caimo} which substitutes iid draws from the likelihood with draws from an auxiliary Markov chain admitting the ERG likelihood as limiting distribution. Previous studies have shown that convergence of sampling from the ERG likelihood through Markov chain is likely to be exponentially slow \citep{bhamidi}. This is likely to lead to increased computational burden when analyzing graphs with complex dependencies.
%%%%%%%%%%%%%%%%%%%%%%%%%%%%%%%%%%%%%%%%%%
\vspace{-1.0em}
\begin{figure}[H]
    \centering
		\includegraphics[width=15cm,height=8cm]{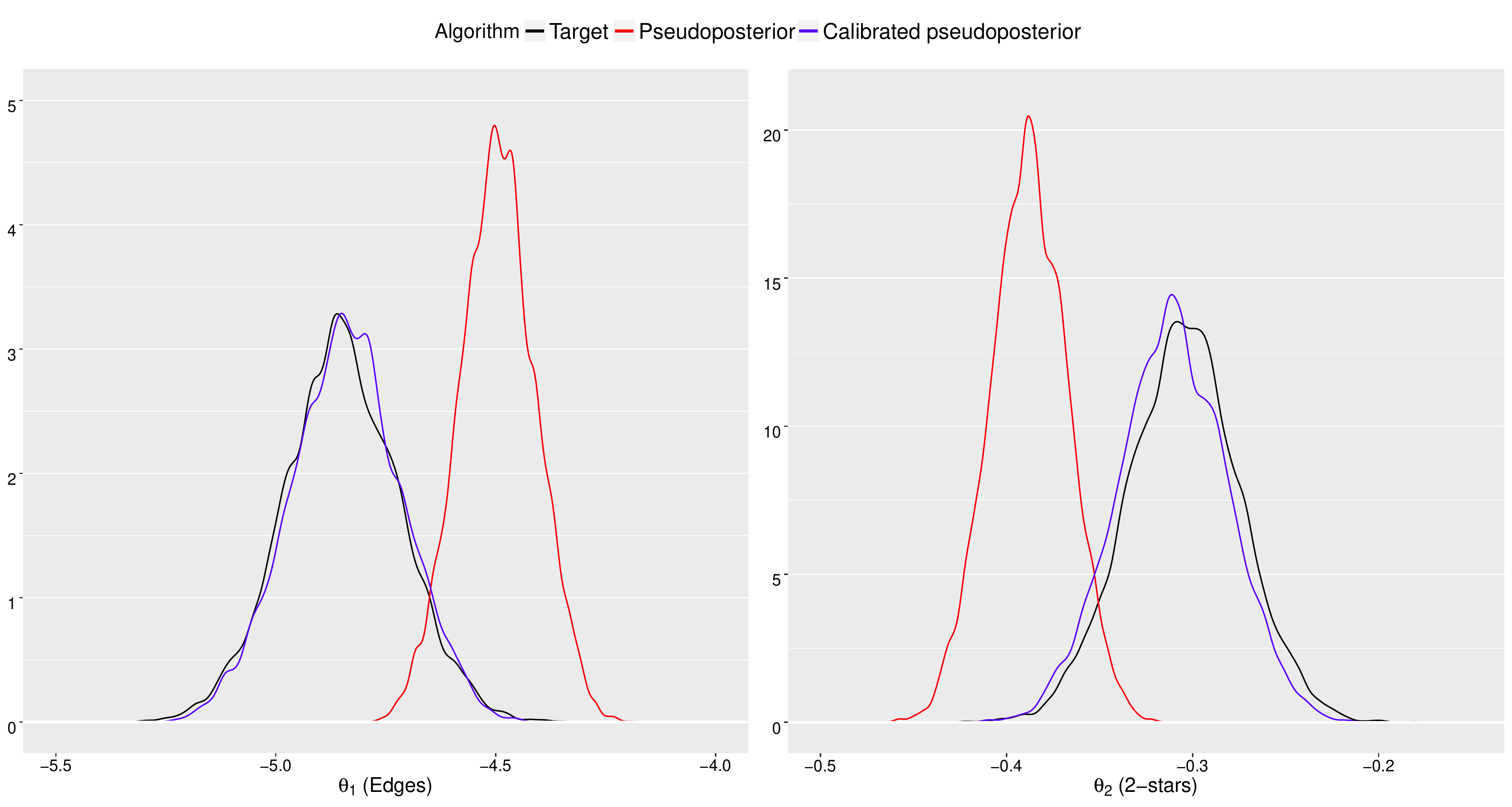}
		\caption{International E--road network: marginal density estimates of the posterior distribution (based on a long and computationally expensive MCMC run) (black curve);
		misspecified pseudo--posterior density estimates (where the pseudolikelihood has replaced the likelihood) (red curve); calibrated pseudo--posterior density estimates (blue curve).
		One can see that the calibration step has resulted in density estimates which are very close to the target posterior.}
\label{fig:euroroad_PL_marginals}
\end{figure}
%%%%%%%%%%%%%%%%%%%%%%%%%%%%%%%%%%%%%%%%%%
\noindent Increasing computational complexity has motivated the development of misspecified but tractable and computationally affordable models. 
From this perspective, the computational tractability of the pseudolikelihood function and the simplicity in defining the objective function seem to make it a tempting alternative to the full likelihood function when dealing with data with such a complex structure \citep{robins,handcock:statnet}. The use of such an approximation to the likelihood should be treated with caution, though, as discussed by \cite{duijn}; their work studied the quality of the pseudolikelihood approximation to the true likelihood, concluding that it may under-estimate endogenous network formation processes. Despite its use by practitioners in the frequentist setting to fit ERG models, little is known about the pseudolikelihood approximation efficiency when embedded in a Bayesian posterior distribution. Our empirical analysis shows that the Bayesian estimators resulting from using the pseudolikelihood function as a plug-in for the true likelihood function are biased and their variance can be underestimated. However, the calibration procedure which we develop shows how to correct a sample from this so-called pseudo-posterior distribution, so that it is approximately distributed from the true posterior distribution.

We propose a novel methodology that falls into the area of MCMC targeting a misspecified posterior: a Metropolis--Hastings sampler whose stationary distribution we refer to as a \textit{pseudo-posterior} (a posterior distribution where the likelihood is replaced by a pseudolikelihood approximation). Such a method is fast and overcomes double intractability but is also {\it{noisy}}, in the sense that it samples from an approximation of the true posterior distribution, $\pi$. A graphical illustration of marginal posterior densities under misspecification can be seen in Fig. \ref{fig:euroroad_PL_marginals}.
This example involves a two--dimensional model which we provide details of in Section \ref{section:euroroad}. Here the misspecification of the actual posterior yields a disastrous approximation; compare the black and red curves. It is evident that a sampler which targets an approximate posterior resulting from replacing the intractable likelihood with a pseudolikelihood approximation leads to biased posterior mean estimates and considerably underestimated posterior variances. Nevertheless, we will present a correction method that allows to calibrate the sample to the true density; warping the red to the blue curve which is now a sensible approximation to the black one.

The aim of this paper is to exploit the use of pseudolikelihoods in Bayesian inference of Exponential random graph models. Our work explores the computational efficiency of the resulting Markov chains, and the trade--off between computational and statistical efficiency in estimates derived from such pseudo--posteriors in comparison to the approximate exchange algorithm of \cite{caimo}. We present the reader with a viable approach to calibrate the posterior distribution resulting from using a misspecified likelihood function in an efficient manner (Fig. \ref{fig:euroroad_PL_marginals}), while providing the theoretical framework for this approach.

The outline of the article is as follows. A basic description of Exponential random graph models is given in Section \ref{section:ergm}. The pseudolikelihood function as a surrogate for the true likelihood is introduced in Section \ref{section:comp_like}. In Section \ref{section:bayes_inf} we formulate the Bayesian model in the presence of likelihood misspecification, and discuss the theoretical properties and practical aspects of the calibration of the posterior distribution. In Section \ref{section:applications}, we illustrate our methods through numerical examples involving real networks. We conclude the paper in Section \ref{section:discussion} with final remarks.

%====================================================================================
%====================================================================================
\section{Exponential Random Graph Models} \label{section:ergm}
Networks are relational data represented as graphs, consisting of nodes and edges. Many probability models have been proposed in order to understand, summarize and forecast the general structure of graphs by utilizing their local properties. Among those, Exponential random graph models play an important role in network analysis since they can represent transitivity and other structural features in network data that define complicated dependence patterns not easily modeled by more basic probability models (\cite{wasserman}, see also \cite{robins2} for a review and the references therein for more details).

Let $\mathcal{Y}$ denote the set of all possible graphs on \textit{n} nodes. The network topology structure is measured by a $n\times n$ random adjacency matrix \textit{Y} of a graph on \textit{n} nodes (actors) and a set of edges (relationships). The presence or absence of a tie between node \textit{i} and \textit{j} is coded as $\textit{Y}_{ij} = 1$ if the dyad (\textit{i}, \textit{j}) is connected, $Y_{ij} = 0$ otherwise. An edge connecting a node to itself is not permitted so $Y_{ii} = 0$.

ERG models are a particular class of discrete linear exponential families which represent the probability distribution of a random network graph using a likelihood function, expressed as:
%%%%%%%%%%%%%%%%%%%%%%%%%%%%%%%%%%
\begin{align} \label{eqn:ergmprob}
p(y\mid \theta)=\frac{q(y\mid \theta)}{z(\theta)}=\frac{\text{exp}\left\{\theta^{\text{T}}s(y)\right\}} {\sum_{y\in \mathcal{Y}}^{} \text{exp}\left\{\theta^{\text{T}}s(y)\right\}},\qquad \theta^{\text{T}}s(y)=\sum_{j=1}^{d}\theta_{j}s_{j}(y)\,,
\end{align}
%%%%%%%%%%%%%%%%%%%%%%%%%%%%%%%%%%
where $q(\text{y}\mid \theta)$ is the unnormalized likelihood. In this paper we deal with ERG models that are edge-dependent, in which case the likelihood is intractable and the pseudolikelihood provides an inaccurate approximation. For such models, $s(y)~=~\{s_1(y),\ldots,s_d(y)\}$ is a known vector of overlapping sub--graph configurations/ sufficient statistics (e.g. the number of edges, degree statistics, triangles, etc.) and $\theta\in\Theta\subseteq\mathbb{R}^{\text{d}}$ are the model parameters. More examples can be found at \cite{snijders_recent_2007} and \cite{hunter}. A positive parameter value for $\theta_i\in\theta$ results in a tendency for the certain configuration corresponding to $s_i(y)$ to be observed in the data than would otherwise be expected by chance.

Despite their popularity, Bayesian parameter estimation for these models is challenging, since evaluation of the posterior typically involves the calculation of an intractable normalizing constant $z(\theta)$. Its evaluation is extremely difficult for all but trivially small graphs since this sum involves $2^{\binom{n}{2}}$ terms for undirected graphs. The intractable normalizing constant makes inference difficult for both frequentist and Bayesian approaches.

%====================================================================================
%====================================================================================
\section{Pseudolikelihood Approximation of Likelihood Function}\label{section:comp_like}
Complex dependencies in several applications lead to full likelihoods that may be difficult, or even impractical, to compute. In such situations it may be useful to resort to approximate likelihoods, if it is possible to compute the likelihood for some subsets of the data.

The pseudolikelihood of \cite{besag1,besag2} was the earliest method of parameter estimation that was proposed for models with complicated dependence structure for which the likelihood function could not be calculated exactly, or even approximated well in a reasonable amount of time. The composite likelihood of \cite{lindsay} is a generalization of the pseudolikelihood (see \cite{varin1} for a recent review).

\cite{strauss} have applied the idea of pseudolikelihood to social networks. The pseudolikelihood method defines the approximation to the full joint distribution as the product of the full conditionals for individual observations/ dyads:
%%%%%%%%%%%%%%%%%%%%%%%%%%%%%%%%
\begin{align}\label{eqn:pseudo}
p^{}_{\text{PL}}(y\mid \theta)=\prod_{i\neq j}^{}p(y_{ij}\mid y_{-ij},\theta)=\prod_{i\neq j}^{}\frac{p(y_{ij}=1\mid y_{-ij},\theta)^{y_{ij}}}{\{1-p(y_{ij}=1\mid y_{-ij},\theta)\}^{y_{ij}-1}},
\end{align}
%%%%%%%%%%%%%%%%%%%%%%%%%%%%%%%%
where $\text{y}_{-ij}$ denotes $\text{y}\backslash\{\text{y}_{ij}\}$. The distribution of the Bernoulli variable $Y_{ij}$, conditional on the rest of the network, can be easily calculated under an alternative specification of model (\ref{eqn:ergmprob}). Let us define the vector of change statistics as
%%%%%%%%%%%%%%%%%%%%%%%%%%%%%%%%%%
\[
\delta_{s}(y)_{ij}=s(y^{+}_{ij})-s(y^{-}_{ij}).
\]
%%%%%%%%%%%%%%%%%%%%%%%%%%%%%%%%%%
This vector is associated with a particular pair (ordered or unordered, depending on whether the network is directed or undirected) of nodes and it equals the change in the vector of network sufficient statistics when that pair is toggled from a 0 (no edge, $y^{-}_{ij}$) to a 1 (edge, $y^{+}_{ij}$), holding the rest of the network, $y_{-ij}=y\backslash\{y_{ij}\}$, fixed. Using such a reparameterization allows to express the distribution of the Bernoulli variable $Y_{ij}$, conditional on the rest of the network, as a function of the change statistics vector:
%%%%%%%%%%%%%%%%%%%%%%%%%%%%%%%%%%
\begin{align}\label{eqn:logitic}
p(y_{ij}=1\mid y_{-ij},\theta)=\text{logit}^{-1}\left\{\theta^{\text{T}}\delta_{s}(y)_{ij}\right\}.
\end{align}
%%%%%%%%%%%%%%%%%%%%%%%%%%%%%%%%%%
This expression is extremely convenient as the values $\{p(y_{ij}=1\mid y_{-ij},\theta)\}_{i\neq j}$, referred to as the predictor matrix, are tractable, fast to compute and require the computation of the change statistics only once up front. The predictor matrix for a new parameter can be updated just by computing the scalar product $\theta^{\text{T}}\delta_{s}(y)$ for the new parameter. As a consequence, estimating the pseudolikelihood $p^{}_{\text{PL}}(y\mid \theta)$ in Eq. \eqref{eqn:pseudo} for any $\theta\in\Theta$ is effortless.

However, the method implicitly relies on the strong and often unrealistic assumption of independent dyads, conditionally on their neighbours. Its properties are poorly understood \citep{duijn}: it does not directly maximize the likelihood and in empirical comparisons \citep{corander} has larger variability than the MLE. In the context of the autologistic distribution in spatial statistics, \cite{friel} showed that the pseudolikelihood estimator can lead to inefficient estimation.

From Eq. \eqref{eqn:pseudo} and \eqref{eqn:logitic}, the pseudolikelihood for model \eqref{eqn:ergmprob} is identical to the likelihood for a logistic regression model. A detailed discussion on the use of the pseudo-likelihood and its relationship with logistic regression is provided by \cite{wasserman}. In that setting, the binary response data consist of the off--diagonal elements of the observed adjacency matrix \citep{cranmer}. Each row of the matrix of change statistics is associated with the vector of change statistics for a particular pair of nodes and can also accommodate for individual attributes and dyadic covariates. This kind of formulation leads to efficient computational algorithms.

%====================================================================================
%====================================================================================
\section{Bayesian Inference for ERGMs}\label{section:bayes_inf}
In Bayesian statistics, the uncertainty about the unknown parameters is modeled by assigning a probability distribution to those parameters of interest that are therefore regarded as random variables. During the last decade there has been an increasing interest for modeling networks under the Bayesian framework. Recently, Bayesian approaches for inferring ERG models have been proposed by \cite{koskinen2}, \cite{caimo3} and \cite{caimo4}. The focus of interest in Bayesian inference is the posterior distribution:
%%%%%%%%%%%%%%%%%%%%%%%%%%%%%%%%%%%%%%%%%%
\begin{align} \label{eqn:postdistr1}
\pi(\theta\mid  y)=\frac{p(y\mid \theta)p(\theta)}{\pi(y)},
\end{align}
%%%%%%%%%%%%%%%%%%%%%%%%%%%%%%%%%%%%%%%%%%
where $p(\theta)$ is a prior distribution on $\theta$. Throughout the article we consider a prior distribution that has a simple form, to ensure that it can be simulated using standard techniques and that the posterior always exists. Markov chain Monte Carlo can be used to infer $\pi(\theta\mid y)$. 

Consider the use of a Metropolis-Hastings (MH) update \citep{metropolis1953equation,hastings}. Given a proposal $h(\theta'\mid \theta)$ the algorithm proposes the move from $\theta$ to $\theta'$ with probability
%%%%%%%%%%%%%%%%%%%%%%%%%%%%%%%%%%%%%%%%%%
\begin{align*}% \label{eqn:MH_accept_prob}
A(\theta,\theta')=\min\left\{1,
\frac{q(y\mid \theta')}{q(y\mid \theta)}
\frac{p(\theta')}{p(\theta)}
\frac{h(\theta\mid \theta')}{h(\theta'\mid \theta)}
\frac{z(\theta)}{z(\theta')}\right\}
\end{align*}
%%%%%%%%%%%%%%%%%%%%%%%%%%%%%%%%%%%%%%%%%%
and has a stationary distribution of $\pi(\theta\mid  y)$. When the likelihood is known in closed form, an exact MCMC targeting $\pi$ can be implemented. However, when a likelihood with a very large number of observations or a likelihood with unknown normalizing constant is present, as is the case for ERGMs, standard MCMC algorithms cannot be implemented.

Being a function of the parameters, the normalizing term $z(\theta)$ cannot be ignored. The acceptance probability is dependent on a ratio of intractable normalizing constants; therefore, a naive application of the Metropolis-Hastings algorithm is not feasible here. This has generated research to find Markov chains that are exact and can be implemented is these situations, such as the exchange algorithm, presented below.

\subsection{The Exchange Algorithm}\label{section:exch_alg}
A popular approach for dealing with intractable likelihoods, such as Markov random fields, is the exchange algorithm. \cite{murray} extended the work of \citet{moller} to allow inference on doubly intractable distributions using the exchange algorithm. The exchange samples from the following augmented distribution
%%%%%%%%%%%%%%%%%%%%%%%%%%%%%%%%%%%%%%%%%%
\begin{align} \label{eqn:postdistr2}
\pi(\theta',y',\theta\mid y)\propto p(y\mid \theta)p(\theta)h(\theta'\mid \theta)p(y'\mid \theta'),
\end{align}
%%%%%%%%%%%%%%%%%%%%%%%%%%%%%%%%%%%%%%%%%%
where $p(y'\mid \theta')$ is the same distribution as the original distribution on which the data $y$ is defined.

This algorithm deals with the intractability of the normalizing constant by introducing an auxiliary variable, hence sampling on an extended state space, that allows the cancellation of all of the normalizing constants in the acceptance probability. This gives rise to a Metropolis--Hastings acceptance probability which is tractable even for target distributions whose normalizing constant depends on the parameter of interest (doubly intractable problems).

The exchange algorithm requires exact simulation of the auxiliary variable $y'$ from the likelihood; perfect sampling from the likelihood is feasible for other MRFs like the Ising and Potts models \citep{propp,huber1}, but this is not the case for ERG models. The {\it{approximate exchange algorithm}} (AEA) of \citet{caimo} modifies the original exchange algorithm and makes it applicable also in settings where sampling from the auxiliary likelihood (the ERG likelihood in our case) is not feasible. It is possible to carry out inference for graphs of larger size (eg. 1000 nodes), but at the cost of an increased computational time.

The approximate exchange resorts to the tie-no-tie (TNT) sampler \citep{ergm} to get approximate draws. The TNT sampler simulates a Markov chain whose transition kernel admits $p(y'\mid \theta')$ as limiting distribution. The Markov kernel is iterated a large number of times, $M$, so that the final point is approximately distributed under $p(y'\mid \theta')$. At this stage it can be anticipated that the number of TNT iterations will be proportional to the number of dyads of the graphs, $n^2$, for a graph with {\textit{n}} nodes. This fact has been theoretically supported by the two following recent studies:
\begin{itemize}
\item \cite{Everitt} has proved that when the MCMC kernel for the exact exchange algorithm is uniformly ergodic, the invariant distribution (when it exists) of the corresponding approximate exchange algorithm becomes closer to the “true” target (that of the exact exchange algorithm) as the number of auxiliary iterations, $M$, increases.
\item \cite{bhamidi} have shown that convergence of sampling from an ERG model through Markov chain Monte Carlo (typically the TNT sampler) is likely to be exponentially slow.  Therefore, this suggests that one should take a conservative approach and choose a large number of auxiliary iterations for a graph with 100 nodes. The resulting exchange algorithm may be infeasible for large graphs due to the exponentially long mixing time for the auxiliary draw from the likelihood.
\end{itemize}

\subsection{Bayesian Pseudo-posteriors}\label{section:bayes_complike}
The aforementioned issues relating to the exchange and approximate exchange algorithms have motivated tractable approximations to the likelihood function which we consider here in the context of Bayesian inference. More specifically, we propose to replace the true likelihood $p(y\mid \theta)$ with a misspecified likelihood model, leading us to focus on the approximated posterior distribution \citep{pauli}, or "pseudo--posterior":
%%%%%%%%%%%%%%%%%%%%%%%%%%%%%%%%%%%%%%%%%%
\begin{align} \label{eqn:postdistr4}
\pi^{}_{\text{PL}}(\theta\mid y)&\propto p^{}_{\text{PL}}(y\mid \theta)p(\theta).
\end{align}
%%%%%%%%%%%%%%%%%%%%%%%%%%%%%%%%%%%%%%%%%%
\noindent To conduct Bayesian inference using Eq. (\ref{eqn:postdistr4}), we adopt the full-update Metropolis-Hastings sampler. The acceptance probability is now tractable but approximated, due to the use of a misspecified model. Algorithm \ref{alg:PL_MhG} provides the pseudocode for the Metropolis--Hastings algorithm that samples from a pseudo-posterior and forms the basis for our correction method.
%%%%%%%%%%%%%%%%%%%%%%%%%%%%%%%%%%%%%%%%%%
\begin{algorithm}[H]
\caption{Pseudolikelihood-based Metropolis–Hastings sampler}
\label{alg:PL_MhG}
\begin{algorithmic}[1]
 \State {\bf{Input:}} $
        \begin{aligned}[t]
				& \text{Initial setting:}~\theta; \\
				& \text{A proposal distribution,}~h(\cdot\mid \theta);\\
        & \text{Number of iterations, T};\\
        & \text{Matrix of change statistics.}\\
				\end{aligned}$
\State {\bf{Output:}} A realization of length T from a Markov chain.	
\For{$\text{t}~=~0,\ldots,\text{T}$}
		\State Propose $\theta'\sim h(\cdot\mid \theta)$;
		\State $\widetilde{A}\big(\theta^{(t)},\theta'\big)~=~\min\bigg\{1, 
		\frac{p^{}_{\text{PL}}\big(y\mid \theta'\big)}{p^{}_{\text{PL}}\big(y\mid \theta^{(t)}\big)}
		\frac{p\big(\theta'\big)}{p\big(\theta^{(t)}\big)}
		\frac{h\big(\theta^{(t)}\mid \theta'\big)}{h\big(\theta'\mid \theta^{(t)}\big)}\bigg\}
		$;
		\State Draw u$~\sim~\text{Uniform}[0,1]$;
		\State {\bf{if}} $\text{u}~\leq~\widetilde{A}\big(\theta^{(t)},\theta'\big)$ {\bf{then}} $\theta^{(t+1)}~\leftarrow~\theta'$;
\EndFor
\State {\bf{return}} $\{\theta_{t}\}_{\text{t}=1,\ldots,\text{T}}$;
\end{algorithmic}
\end{algorithm}
%%%%%%%%%%%%%%%%%%%%%%%%%%%%%%%%%%%%%%%%%%

\subsection{Adjustment of the pseudo--posterior distribution}\label{section:calibration}
Recent work by \cite{friel3} focused on calibrating conditional composite likelihoods with overlapping components/ blocks. They observed that a non--calibrated composite likelihood (the generalization of the pseudolikelihood misspecification) leads to an approximated posterior distribution with substantially lower variability than the true posterior distribution, leading in turn to overprecision about posterior parameters.

Calibration of the unadjusted posterior sample to obtain appropriate inference, as described in their work, is executed with two operations: a "mode adjustment" to ensure that the true and the approximated posterior distributions have the same mode and a "curvature adjustment" that modifies the geometry of the approximated posterior at the mode.

We will use the following notation:
%%%%%%%%%%%%%%%%%%%%%%%%%%%%%%%%%%%%%%%%%%
\begin{align*}
\arg\max_{\theta}\pi(\theta\mid y)=\theta^{\ast}&,~\Hcal \log\pi(\theta\mid y)|_{\theta^{\ast}}=H^{\ast},\\
\arg\max_{\theta}\pi^{}_{\text{PL}}(\theta\mid y)=\htheta_{\text{PL}}&,~\Hcal \log\pi^{}_{\text{PL}}(\theta\mid y)|_{\htheta_{\text{PL}}}=\hat{H}^{}_{\text{PL}},
\end{align*}
%%%%%%%%%%%%%%%%%%%%%%%%%%%%%%%%%%%%%%%%%%
where for any function $f$ taking values in the set of strictly positive real numbers and assuming that $\log f$ is twice differentiable at $\theta_0$, $\Hcal\log f(\theta)|_{\theta_0}$ is the Hessian matrix of $\log f$ at $\theta_0$. In the ERGM context the parameters $\theta^{\ast}$, $H^{\ast}$, $\htheta_{\text{PL}}$ and $\hat{H}^{}_{\text{PL}}$ are either available in closed form or can be estimated using Monte Carlo. In the second half of this section we provide practical guidelines on how to obtain these. We denote by $\widetilde{\pi}^{}(\theta\mid y)$ the fully calibrated target
whose mode is located at $\thetas$ and whose second order derivative at the mode is equal to $H^{\ast}$:
%%%%%%%%%%%%%%%%%%%%%%%%%%%%%%%%%%%%%%%%%%
\begin{align}\label{eq1}
\arg\max_{\theta}\widetilde{\pi}^{}(\theta\mid y)=\theta^{\ast},~\Hcal\log\widetilde{\pi}^{}(\theta\mid y)|_{\theta^{\ast}}={H}^{\ast}.
\end{align}
%%%%%%%%%%%%%%%%%%%%%%%%%%%%%%%%%%%%%%%%%%
\noindent
In this paper, we choose to adjust the pseudo--posterior through an affine transformation $\text{g}(\theta)=W\theta+\lambda$, where $W\in\mathcal{M}(\rset^{d})$ is an invertible matrix. More precisely, the calibrated pseudo--posterior has the following probability density function:
%%%%%%%%%%%%%%%%%%%%%%%%%%%%%%%%%%%%%%%%%%
\begin{equation}\label{eq2}
\widetilde{\pi}^{}(\theta\mid y)\propto\pi^{}_{\text{PL}}[\text{g}(\theta)\mid y]\,.
\end{equation}
%%%%%%%%%%%%%%%%%%%%%%%%%%%%%%%%%%%%%%%%%%
\noindent To satisfy the two equations in \eqref{eq1}, we identify $W$ and $\lambda$ as follows:
\\\newline
\noindent \textbf{Condition 1:}
%%%%%%%%%%%%%%%%%%%%%%%%%%%%%%%%%%%%%%%%%%
\begin{equation*}
\left. \nabla_{\theta}\log\widetilde{\pi}^{}(\theta\mid y) \right|_{\thetas}=0\,,
\end{equation*}
%%%%%%%%%%%%%%%%%%%%%%%%%%%%%%%%%%%%%%%%%%
which, irrespective of the choice of $W$, gives
%%%%%%%%%%%%%%%%%%%%%%%%%%%%%%%%%%%%%%%%%%
\begin{eqnarray*}
    \left.\nabla_{\theta}\log\pi^{}_{\text{PL}}(\theta\mid y) \right|_{\htheta_{\text{PL}}} &=& 0  \\
    \iff  \nabla_{\theta}\log\widetilde{\pi}^{}(\theta\mid y)&=& \left.W^T\nabla_{\theta}\log\pi^{}_{\text{PL}}(\theta\mid y)\right|_{W\theta+\lambda}  \\
    \Rightarrow \lambda &=& \htheta_{\text{PL}}-W\thetas.
\end{eqnarray*}
%%%%%%%%%%%%%%%%%%%%%%%%%%%%%%%%%%%%%%%%%%
\noindent \textbf{Condition 2:}
%%%%%%%%%%%%%%%%%%%%%%%%%%%%%%%%%%%%%%%%%%
\begin{equation}\label{eq3}
\left.\Hcal\log\widetilde{\pi}^{}(\theta\mid y)\right|_{\thetas}=H^{\ast},
\end{equation}
%%%%%%%%%%%%%%%%%%%%%%%%%%%%%%%%%%%%%%%%%%
where the left--hand side can be written as
$$
\left.\Hcal\log\widetilde{\pi}^{}(\theta\mid y)\right|_{\thetas}=W^T\left.\Hcal\log\pi^{}_{\text{PL}}(\theta\mid y)\right|_{W(\thetas-\thetas)+\htheta_{\text{PL}}}W=W^T\hat{H}^{}_{\text{PL}}W.
$$
Since $H^{\ast}$ and $\hat{H}^{}_{\text{PL}}$ are both Hessians at the mode of their respective distribution, they are negative-definite matrices and therefore admit a Cholesky decomposition: $-H^{\ast}=N^T N$ and $-\hat{H}^{}_{\text{PL}}=M^T M$. Based on this observation, it is straightforward to check that the choice $W=M^{-1}N$ satisfies Eq. \eqref{eq3}. $\widetilde{\pi}^{}$ is now a proper density entirely specified by
%%%%%%%%%%%%%%%%%%%%%%%%%%%%%%%%%%%%%%%%%%
\begin{align}\label{eq4}
\widetilde{\pi}^{}(\theta\mid y)=|\text{det}(W)| \;\pi^{}_{\text{PL}}[W(\theta-\thetas)+\htheta_{\text{PL}}\mid y]\,
\end{align}
%%%%%%%%%%%%%%%%%%%%%%%%%%%%%%%%%%%%%%%%%%
and satisfies the two conditions $\arg\max_{\theta}\widetilde{\pi}^{}(\theta\mid y)=\theta^{\ast}$ and $\Hcal\log\widetilde{\pi}^{}(\theta\mid y)|_{\theta^{\ast}}={H}^{\ast}$.

Section \ref{section:bayes_complike} shows how one can gather a sample $(\theta_1,\ldots,\theta_T)$ from $\pi^{}_{\text{PL}}$. The correction step then consists of applying the mapping $g^{-1}:\Theta\to\Theta$ so that the corrected sample
%%%%%%%%%%%%%%%%%%%%%%%%%%%%%%%%%%%%%%%%%%
\begin{align*}
\zeta_i=g^{-1}(\theta_i)=V\theta_i+(\thetas-V\htheta_{\text{PL}}),~\text{where}~V=W^{-1},
\end{align*}
%%%%%%%%%%%%%%%%%%%%%%%%%%%%%%%%%%%%%%%%%%
for $i=1,\ldots,T$, is distributed under $\widetilde{\pi}^{}$.

\subsection{Optimization algorithms}\label{section:optimization}
It is clear that implementing such a transformation is feasible provided that the quantities $V$, $\hat{\theta}^{*}$ and $\htheta_{\text{PL}}$ are available. Equivalently, once those quantities are estimated from the data and without having performed any MCMC sampling in advance as described in section \ref{section:bayes_complike}, one can opt for sampling directly from the fully calibrated pseudo-posterior distribution, whose analytical form is known (Eq. \ref{eq4}), with the Metropolis--Hastings algorithm. We now provide practical guidelines for estimation of the parameters $\theta^{\ast}$, $\htheta_{\text{PL}}$, $H^{\ast}$ and $\hat{H}^{}_{\text{PL}}$.

\begin{enumerate}
\item To estimate $\theta^{*}_{\text{PL}}$ one can use any gradient-based optimiser. Here we used a BFGS algorithm \citep{nocedal} which was based on available exact gradient evaluations. In practice, the gradient evaluated in the
BFGS algorithm is calculated using standard logistic regression theory \citep{hosmer}.

\item The Robbins--Monro stochastic approximation algorithm \citep{robbins} can be used to estimate the maximum a posteriori $\theta^{*}$, when the gradient of the (log)posterior is not analytically tractable. The Robbins--Monro algorithm takes steps in the direction of the slope of the distribution and follows the stochastic process
    $$
    \theta_{i+1}=\theta_{i}+\epsilon_{i}\left.\hat{\nabla}_{\theta}\log{\pi(\theta_i\mid y)}\right.\,,
    \qquad\text{where}\quad\sum\limits_{i=0}^{\text{N}}\epsilon_i=\infinity\;\text{and}\;\sum\limits_{\text{i}=0}^{\text{N}}\epsilon_i^2<\infinity\,,
    $$
    where $\hat{\nabla}_{\theta}\log{\pi(\theta_i\mid y)}$ denotes a noisy version of the gradient of the log-posterior at $\theta_i$ and $\{\epsilon_i\}_{i}$ is a non-increasing sequence of positive numbers. In our analysis we use a sequence of steps which satisfy these conditions, having the form $\epsilon_i=\alpha/i,~\alpha>0$. Once the difference between successive values of this process is less than a specified tolerance level, the algorithm is deemed to have converged to the MAP. The noisy gradient of the log-posterior at $\theta$ is derived as follows. First, the logarithm of the posterior can be written as:
%%%%%%%%%%%%%%%%%%%%%%%%%%%%%%%%%%%%%%%%%%
\[
\log{\pi(\theta\mid y)}=\theta^{\text{T}}s(y)-\log{z(\theta)}+\log{p(\theta)}-\log{\pi({y)}}.
\]
%%%%%%%%%%%%%%%%%%%%%%%%%%%%%%%%%%%%%%%%%%
\noindent The first derivative of the log-posterior with respect to $\theta$ yields:
%%%%%%%%%%%%%%%%%%%%%%%%%%%%%%%%%%%%%%%%%%
\begin{align} \label{eqn:first_deriv_true_post}
\nabla_{\theta}\log{\pi(\theta\mid y)}
&=s(y)-\sum_{y\in \mathcal{Y}}^{}s(y)p(y\mid \theta)+\nabla_{\theta}\log{p(\theta)}\\\nonumber
&=s(y)-\mathbb{E}_{\text{y}\mid \theta}\left[s(y)\right]+\nabla_{\theta}\log{p(\theta)}.
\end{align}
%%%%%%%%%%%%%%%%%%%%%%%%%%%%%%%%%%%%%%%%%%
\noindent The exact evaluation of $\mathbb{E}_{\text{y}\mid \theta}\left[s(y)\right]$ is intractable in practice due to the presence of the normalizing term. However it is possible to estimate
$\mathbb{E}_{\text{y}\mid \theta}\left[s(y)\right]$ using Monte Carlo sampling. We can simulate $\text{y}_{\theta}=(\text{y}^\prime_1, \text{y}^\prime_2,\ldots,\text{y}^\prime_{\text{N}})\sim p(\cdot\mid \theta)$
and then estimate $\mathbb{E}_{\text{y}\mid \theta}\left[s(y)\right]$ as $\dfrac{1}{\text{N}}\sum\limits_{\text{i}=1}^{\text{N}}s(\text{y}^\prime_{\text{i}})$. Hence the noisy version of the gradient at $\theta$ will be given by:
%%%%%%%%%%%%%%%%%%%%%%%%%%%%%%%%%%%%%%%%%%
\begin{align}\label{eqn:est_first_deriv}
\hat{\nabla}_{\theta}\log{\pi(\theta\mid y)}=s(y)-\frac{1}{\text{N}}\sum_{\text{i}=1}^{\text{N}}s(\text{y}^\prime_{\text{i}})+\nabla_{\theta}\log{p(\theta)}.
\end{align} 
%%%%%%%%%%%%%%%%%%%%%%%%%%%%%%%%%%%%%%%%%%
In all our examples we started the algorithm from the corresponding maximum pseudolikelihood estimate, $\hat{\theta}_{\text{MPLE}}$ and as we explain in Section~\ref{section:toy_example} it will be important to monitor that $\hat{\theta}_{\text{MPLE}}$ does not lie in a degenerate region. 
The Robbins--Monro algorithm is highly sensitive to the starting point in terms of time to convergence; in our context we set the sequence 
$\{\epsilon_i\}_{i}$ small enough by setting $\alpha=0.001$ and N appropriately large to try to avoid this problem. More precisely, the simulation step (TNT sampler) to draw $y'$ was run for 1,000 iterations followed by an extra 12,000 iterations thinned by a factor of 30, yielding $\text{N} = 400$ graphs.

\item The Hessian matrix $\hat{H}^{}_{\text{PL}}$ of the approximated posterior distribution at the mode $\hat{\theta}_{\text{PL}}$ is analytically available.
\item The curvature of the true posterior distribution $H^{\ast}$ at the MAP $\theta^{\ast}$ can be calculated by taking the derivative of Eq. (\ref{eqn:first_deriv_true_post}):
%%%%%%%%%%%%%%%%%%%%%%%%%%%%%%%%%%%%%%%%%%
\begin{align} \label{eqn:second_deriv_true_post}
\mathcal{H}\log{\pi(\theta\mid y)}&=-\frac{z^{\prime\prime}(\theta)z(\theta)-z'(\theta)z'(\theta)}{z^2(\theta)}+\mathcal{H}\log{p(\theta)}\\\nonumber
&=-\left[\frac{z^{\prime\prime}(\theta)}{z(\theta)}-\left(\frac{z'(\theta)}{z(\theta)}\right)^2\right]+\mathcal{H}\log{p(\theta)}\\\nonumber
&=-\bigg\{\mathbb{E}^2_{\text{y}\mid \theta}\left[s(y)\right]-\left[\mathbb{E}_{\text{y}\mid \theta}\left[s(y)\right]\right]^2\bigg\}+\mathcal{H}\log{p(\theta)}\\\nonumber
&=-\mathbb{V}_{\text{y}\mid \theta}\left[s(y)\right]+\mathcal{H}\log{p(\theta)}.
\end{align}
%%%%%%%%%%%%%%%%%%%%%%%%%%%%%%%%%%%%%%%%%%
The Monte Carlo estimators of the expected value $\mathbb{E}_{\text{y}\mid \hat{\theta}^{*}}\left[s(y)\right]$ and the covariance matrix $\mathbb{V}_{\text{y}\mid \hat{\theta}^{*}}\left[s(y)\right]$ are based on a number of random samples of networks drawn from the specified model. With the aforementioned procedure Monte Carlo error emerges due to the simulation approximation of $\mathbb{E}_{\text{y}\mid \theta}\left[s(y)\right]$ and $\mathbb{V}_{\text{y}\mid \theta}\left[s(y)\right]$.
\end{enumerate}
%\vspace{-1.0em}
%%%%%%%%%%%%%%%%%%%%%%%%%%%%%%%%%%%%%%%%%%%
%\begin{algorithm}[H]
%\caption{Calibration of pseudo--posterior draws}
%\label{alg:calib_PL}
%\begin{algorithmic}[1]
  %\State {\bf{Input:}} Unadjusted pseudo--posterior draws, $\theta_t,~\text{t}=1,\ldots,\text{T}$.
	%\State {\bf{Output:}} Mode and curvature--adjusted pseudo--posterior sample, $\zeta_t,~\text{t}=1,\ldots,\text{T}$.
	%\item[]	
  %\item[]	\textbf{MAP estimation}
  %\vspace{.10cm}
	%\State Estimate $\htheta_{\text{PL}}$ (BFGS algorithm) based on exact gradient evaluations
	%\item[]	(logistic regression theory);
  %\State Estimate $\hat{\theta}^{*}$ (Robbins--Monro algorithm) based on a Monte Carlo estimator of $\nabla_{\theta}\log{\pi(\theta\mid y)}$;\vspace{.10cm}
	%\item[]	
  %\item[]	\textbf{Curvature Adjustment}
%\vspace{.10cm}
 %\State Estimate $\hat{H}^{}_{\text{PL}}$ using logistic regression theory;
 %\State Estimate $H^{\ast}$ based on a Monte Carlo estimator of $\mathcal{H}\log{\pi(\theta\mid y)}$;
 %\State Perform Cholesky decompositions of $H^{\ast}$, $\hat{H}^{}_{\text{PL}}$: $-H^{\ast}=N^T N$, $-\hat{H}^{}_{\text{PL}}=M^T M$;
 %\State Calculate $W=M^{-1}N$;
 %\State Calculate the transformation matrix $V$ by inverting W, $V=W^{-1}$;\vspace{.2cm}
%\State {\bf{return}} Adjusted sample $\zeta_t=V(\theta_t-\htheta_{\text{PL}})+\thetas,~\text{t}=1,\ldots,\text{T}$.
%\end{algorithmic}
%\end{algorithm}
%%%%%%%%%%%%%%%%%%%%%%%%%%%%%%%%%%%%%%%%%%%%
\noindent Algorithm \ref{alg:calib_PL} summarizes all the actions that need to be taken to calibrate the posterior draws from the misspecified target distribution.

%====================================================================================
%====================================================================================
\section{Applications}\label{section:applications}
This section demonstrates the performance of Bayesian parameter estimation of Exponential random graph models with the pseudolikelihood--based Metropolis--Hastings algorithm pre-- and post-- calibration of the respective posterior sample and provides comparisons to the approximate exchange algorithm. A toy example and two real networks of increased complexity drawn from disparate fields will be illustrated. In particular, we considered undirected and unweighted graphs. All computations in this paper were carried out with the statistical environment R \citep{rsoft} on a laptop computer with an Intel \textregistered Core\textsuperscript{TM} i7-4500U CPU (1.80GHz) and 16GB RAM.

The MCMC algorithms in this paper sample from the corresponding target distributions with the use of a full--update Metropolis-Hastings sampler. The main MCMC chain consists of 40,000 iterations with a burn--in period of 10,000 iterations. Throughout our analysis we assumed a diffuse Multivariate Normal prior distribution for the model parameters, $\mathcal{MVN}\left(0_{d},30\times I_d\right)$, where $0_{d}$ is the null vector and $I_{d}$ is the identity matrix of size equal to the number of model dimensions d, unless stated otherwise. A conjugate prior is unavailable in the context of an ERG likelihood because of the intractable normalising constant. The {\bf{Bergm}} package for R \citep{caimo2} implements the methodology developed in this paper.
%%%%%%%%%%%%%%%%%%%%%%%%%%%%%%%%%%%%%%%%%%
\begin{algorithm}[H]
\caption{Calibration of pseudo--posterior draws}
\label{alg:calib_PL}
\begin{algorithmic}[1]
  \State {\bf{Input:}} Unadjusted pseudo--posterior draws, $\theta_t,~\text{t}=1,\ldots,\text{T}$.
	\State {\bf{Output:}} Mode and curvature--adjusted pseudo--posterior sample, $\zeta_t,~\text{t}=1,\ldots,\text{T}$.
	\item[]	
  \item[]	\textbf{MAP estimation}
  \vspace{.10cm}
	\State Estimate $\htheta_{\text{PL}}$ (BFGS algorithm) based on exact gradient evaluations
	\item[]	(logistic regression theory);
  \State Estimate $\hat{\theta}^{*}$ (Robbins--Monro algorithm) based on a Monte Carlo estimator of $\nabla_{\theta}\log{\pi(\theta\mid y)}$;\vspace{.10cm}
	\item[]	
  \item[]	\textbf{Curvature Adjustment}
\vspace{.10cm}
 \State Estimate $\hat{H}^{}_{\text{PL}}$ using logistic regression theory;
 \State Estimate $H^{\ast}$ based on a Monte Carlo estimator of $\mathcal{H}\log{\pi(\theta\mid y)}$;
 \State Perform Cholesky decompositions of $H^{\ast}$, $\hat{H}^{}_{\text{PL}}$: $-H^{\ast}=N^T N$, $-\hat{H}^{}_{\text{PL}}=M^T M$;
 \State Calculate $W=M^{-1}N$;
 \State Calculate the transformation matrix $V$ by inverting W, $V=W^{-1}$;\vspace{.2cm}
\State {\bf{return}} Adjusted sample $\zeta_t=V(\theta_t-\htheta_{\text{PL}})+\thetas,~\text{t}=1,\ldots,\text{T}$.
\end{algorithmic}
\end{algorithm}
%%%%%%%%%%%%%%%%%%%%%%%%%%%%%%%%%%%%%%%%%%%

\subsection{Toy example with degeneracy check}\label{section:toy_example}
Here we illustrate using a simulated toy example that the correction procedure embedded in our algorithm does not inadvertently mask degeneracy in the model specification. To this end, we considered a 
Markov model with edge and triangle parameters which is known to be degenerate \citep{snijders_recent_2007}.

\begin{figure}[H]
\centering
\includegraphics[width=8cm,height=8cm]{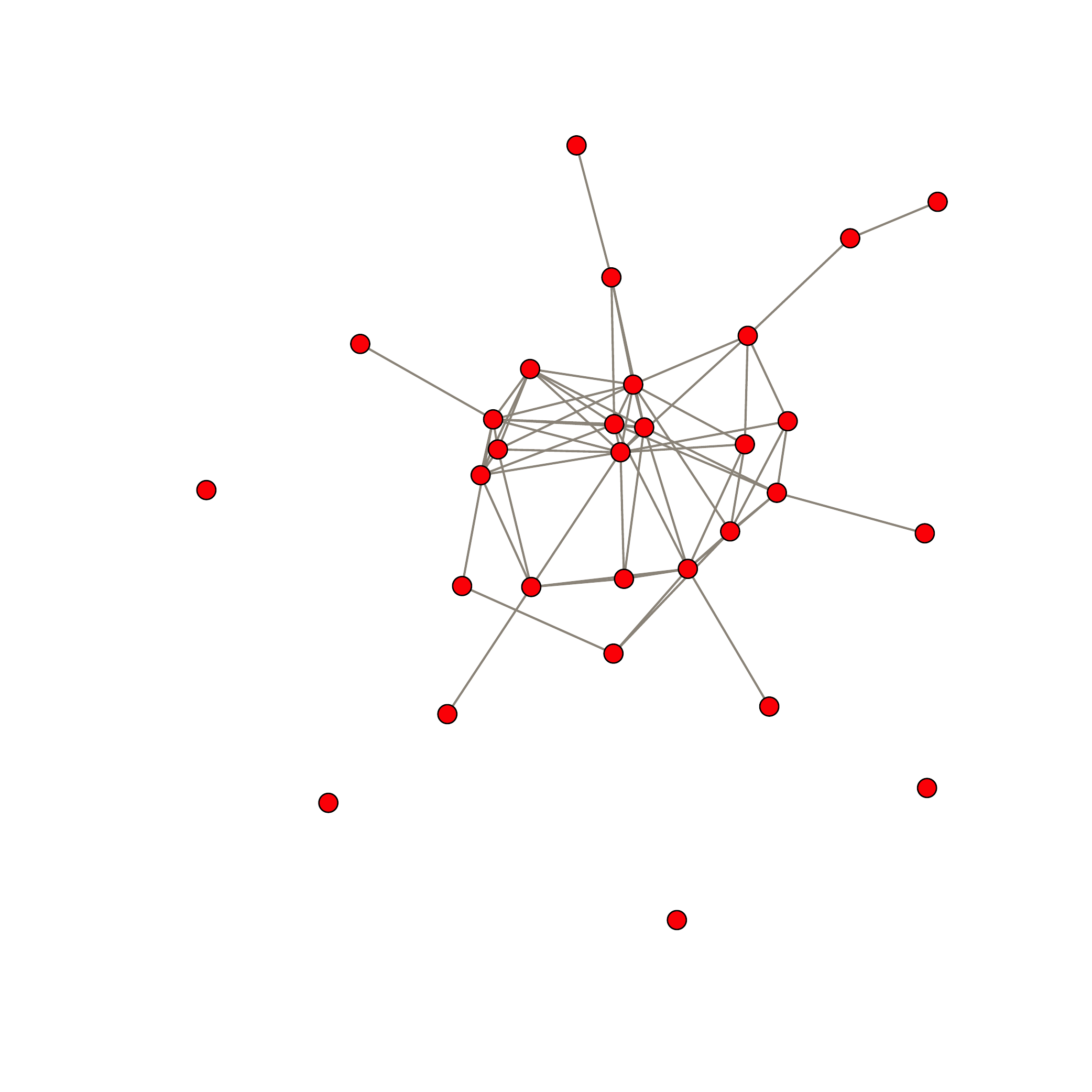}
\vspace{-2.6em}
\caption{Toy example network with 30 nodes and 65 edges.}
\label{fig:toy_example_graph}
\end{figure}

\begin{figure}[H]
\centering
\includegraphics[width=9cm,height=8.8cm]{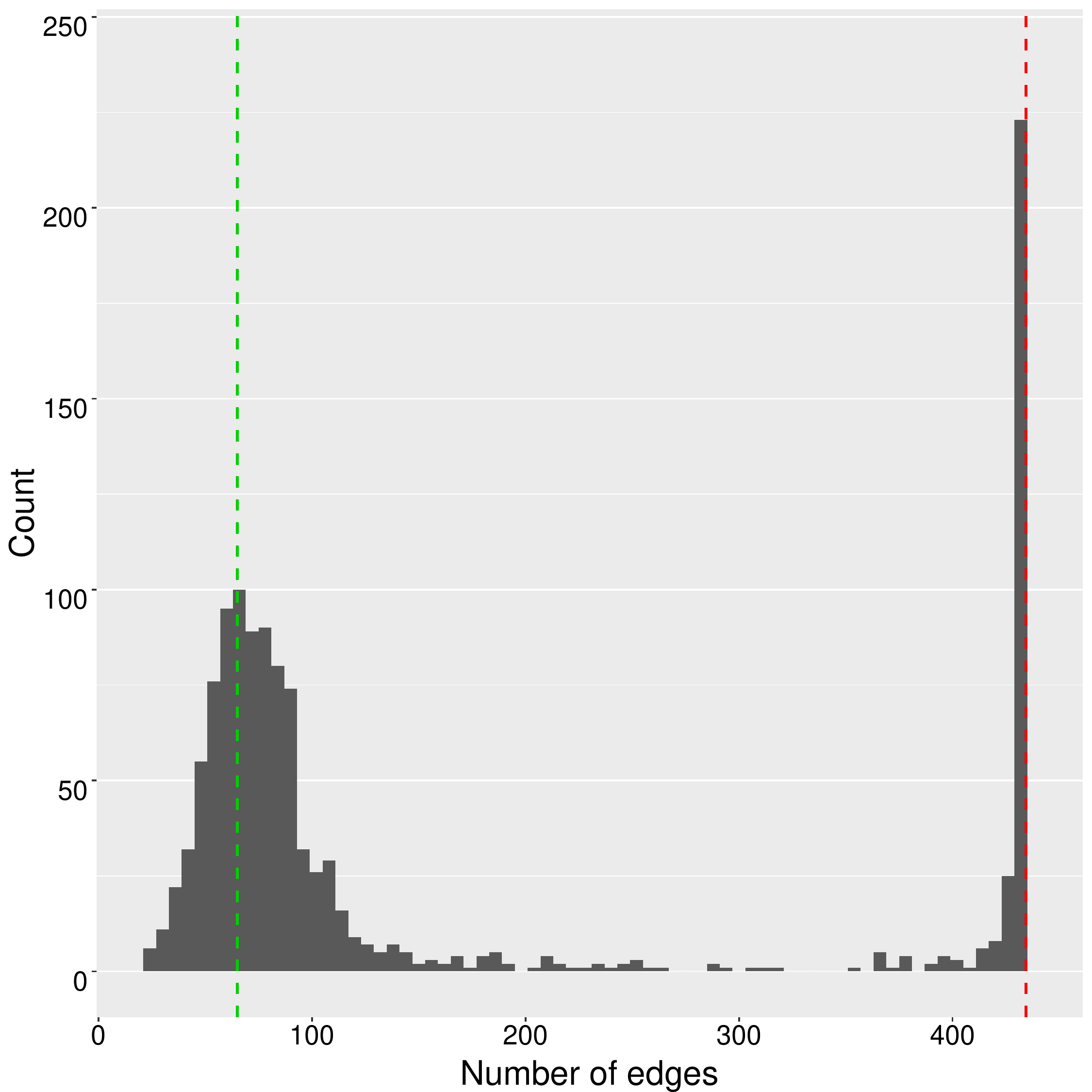}
\vspace{-0.8em}
\caption{Toy example network - Distribution of the number of edges based on 600 simulated graphs from the corrected pseudo-posterior distribution $\widetilde{\pi}^{}(\theta\mid y)$. The green dashed vertical line corresponds to the observed number of edges, while the red dashed vertical line corresponds to the average number of edges based on 20 simulated graphs under the parameter configuration $(\theta^{}_{Edges},\theta^{}_{Triangle})=(-3.0,1.2)$.}
\label{fig:toy_example_graph_degen_check}
\end{figure}

We simulated an undirected graph with 30 nodes with the edge parameter set to -3.0 and triangle parameter set to 1.2 (Fig. \ref{fig:toy_example_graph}). Following the experiments of \cite{snijders_recent_2007}, we expect higher density graphs to arise from simulations from the likelihood under this parameter configuration. We first note that the MPLE was initially used as a starting point for the Robbins–Monro stochastic algorithm. However, since $\hat{\theta}_{MPLE}=(-3.08, 0.95)$ lies in the 
degenerate region (Fig. \ref{fig:toy_example_graph_degen_check}), this causes the algorithm not to converge, as networks simulated from the likelihood around this parameter
vector are typically fully connected and this indeed gives a first indication that this model is inappropriate. As an alternative, we used the $0_{d}$ vector as a starting point and we ran 
the algorithm for 5,000 iterations, where at each iteration 400 graphs were randomly drawn to ensure convergence, see Eq. \eqref{eqn:est_first_deriv}.
%%%%%%%%%%%%%%%%%%%%%%%%%%%%%%%%%%%%%%%%%%
\vspace{-1.1em}
%%%%%%%%%%%%%%%%%%%%%%%%%%%%%%%%%%%%%%%%%%
To check for issues of degeneracy we used the Bayesian procedure of \cite{caimo2}. In particular, we gathered a sample from the corrected pseudo-posterior distribution and simulated  
networks from the likelihood for each $\theta$ parameter in the sample. Note that the mode of the corrected pseudo-posterior distribution was found to be $(\theta^{}_{Edges},\theta^{}_{Triangle})=(-1.895,0.335)$, 
illustrating that the prior distribution, centered at $0_{d}$, has the effect of shifting the high posterior density region away from the degenerate region. This mode is consistent with the estimated mode of 
the true posterior based on a long approximate exchange MCMC run. It is therefore important to check if the posterior distribution still supports parameters in the degenerate region.

To examine this, we generated a sample of size $600$ from $\widetilde{\pi}^{}(\theta\mid y)$, yielding $600$ networks simulated from each parameter in the sample. Fig. \ref{fig:toy_example_graph_degen_check} illustrates the distribution of the number of edges based on those simulated networks. While a reasonable proportion of the simulated networks has a density 
close to the observed network density, there is a large number of networks which are highly connected. This illustrates that the posterior distribution supports some parameter values which lie in the degenerate region, illustrating the unsuitability of this model and reinforcing the need for a model degeneracy check.

\subsection{International E--road Network}\label{section:euroroad}
\begin{figure}[H]
\centering
\includegraphics[width=9cm,height=9cm]{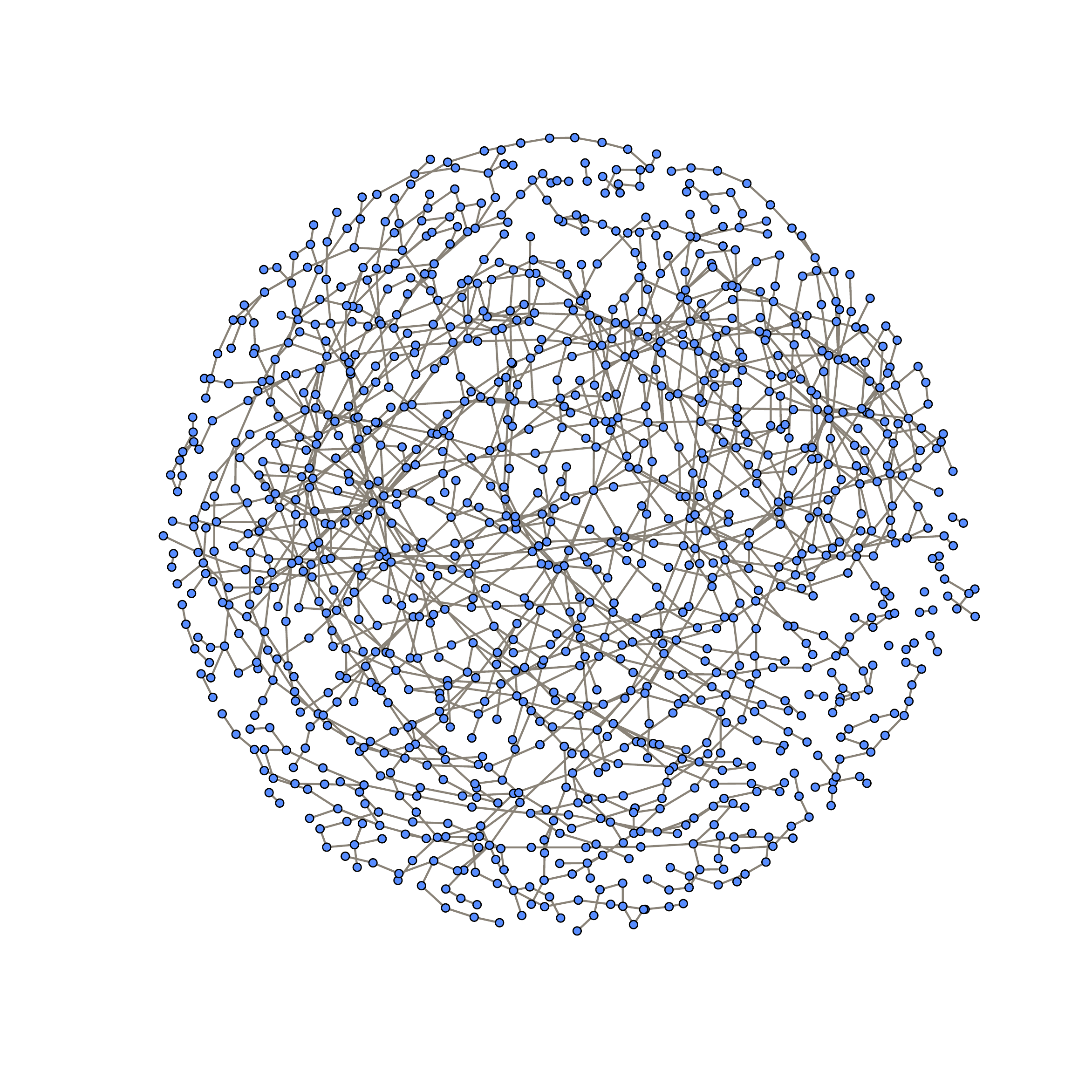}
\vspace{-2.0em}
\caption{International E--road network graph.}
\label{fig:road_graph}
\end{figure}

\noindent The International E--road Network constructed by \cite{subelj} and displayed in Fig. \ref{fig:road_graph} represents all roads included in the International E--road Network. Nodes correspond to European cities and edges represent direct (class A, B) road connections among them. The network has 1177 nodes and 1417 edges. To demonstrate our methodology we fit a 2--dimensional model with edge and 2-star terms, the posterior distribution of which is:
%%%%%%%%%%%%%%%%%%%%%%%%%%%%%%%%%%%%%%%%%%
\begin{align} \label{eqn:postdistr_exchange_euroroad}
\pi(\theta\mid y) \propto \frac{1}{z({\theta)}} \text{exp}\bigg[\theta_{1}\sum_{i<j}^{}y_{ij} + \theta_{2}\sum_{i<j<k}^{}y_{ik}y_{jk}\bigg] p(\theta),
\end{align}
%%%%%%%%%%%%%%%%%%%%%%%%%%%%%%%%%%%%%%%%%%
where $p(\theta)$ is a $\mathcal{MVN}\left(0_{d},30\times I_d\right)$ distribution. In further simulation experiments not shown here, changing the prior distribution to $\mathcal{MVN}\left(0_{d},300\times I_d\right)$ did not lead to different results. We note that the purpose of this example is mainly to illustrate the gains that can result from using our methodology. Such Markov models do not provide a good specification for many real human social networks, where degeneracy and phase transition issues are likely to occur (see \cite{handcock2} and \cite{park} for detailed studies).
%%%%%%%%%%%%%%%%%%%%%%%%%%%%%%%%%%%%%%%%%%
%\vspace{-1.1em}
%%%%%%%%%%%%%%%%%%%%%%%%%%%%%%%%%%%%%%%%%%

\subsubsection{Sampling from the Approximate Exchange}\label{section:euroroad_AEA}
The {\textit{approximate exchange algorithm}} (AEA) was first used to generate draws from the target posterior. The results from this served as a ground truth against which the posterior estimates of $\theta$ using the various approximated likelihood estimators were compared. The {\bf{Bergm}} package \citep{caimo2} for R allows to carry out inference with the approximate exchange algorithm described above.

Bergm uses a multivariate Gaussian proposal distribution. To account for possible correlations between the model parameters, all samplers in this paper assumed the same proposal distribution with a proposal variance-covariance in the form $\Sigma = \text{T}(\text{B}_0 + \text{C}^{-1})^{-1} \text{T}$. T denotes the diagonal positive definite matrix formed from a Metropolis tuning parameter, chosen by the user to reach a reasonable mixing rate, $\text{B}_0$ is the prior precision, and C is the large sample variance--covariance matrix of the MPLEs. The precision matrix $\text{C}^{-1}$ is the same as the negative Hessian, $-H(\hat{\theta})$.

The approximate exchange algorithm requires a number of iterations, $M$, to simulate approximately from the likelihood. A "conservative" approach would be to choose a large number of auxiliary iterations, eg. 500,000, to ensure that the invariant distribution of the approximate exchange algorithm will be considered as the "true" target. Table \ref{tab:euroroad_AEA_tvd} suggests that 10,000 is a practical number of auxiliary iterations for this two-dimensional model.
%%%%%%%%%%%%%%%%%%%%%%%%%%%%%%%%%%%%%%%%%%
\vspace{-1.1em}
%%%%%%%%%%%%%%%%%%%%%%%%%%%%%%%%%%%%%%%%%%
The total variation distance metric \citep{dudley1968} between a given Markov chain and the conservative AEA sample provides a numerical tool to assess the improvement at each step of the calibration. It gives a measure of how well the posterior densities match the "ground truth" posterior density. Recall that the TV--distance of two densities $f(\theta)$ and $g(\theta)$ is given by
\[
TV(f,g) = \frac 1 2 \int\left|f(\theta)-g(\theta)\right|  \operatorname{d}\theta.
\]
\noindent It is equal to 1 when \textit{f} and \textit{g} have disjoint supports and it vanishes when the functions are identical. For the two--dimensional example considered here, the total variation distance was approximated by splitting the lattice into bins with a pre--defined window size. Within each grid the absolute difference of the frequencies was calculated, such that TV takes values in [0,1].
%%%%%%%%%%%%%%%%%%%%%%%%%%%%%%%%%%%%%%%%%%
\begin{table}[H]
\caption{International E--road network: average total variation distance on 20 simulations for Model (\ref{eqn:postdistr_exchange_euroroad}), using the approximate exchange algorithm under an increasing number of auxiliary iterations. The total variation distance was calculated between a given Markov chain and the conservative AEA sample using 500,000 auxiliary iterations.}
\vspace{0.5em}
\centering
\begin{tabular}{lrrrrrr}
\toprule
Auxiliary Iterations          &50  & 100 &300 &500&$10^3$  &$3\times10^3$  \\
\hline
Average TV                    &0.744&0.665&0.603&0.598&0.545&0.234\\
SE of mean ($\times10^{-4}$)  &25&20&28&30&27&21\\
\hline
\hline
\noalign{\vskip 0.02in}
Auxiliary Iterations          &$5\times10^3$ & $10^4$ & $2\times10^4$ & $4\times10^4$ & $10^5$ \\
\hline
Average TV                    &0.075&0.029&0.029&0.030&0.028\\
SE of mean ($\times10^{-4}$)  &19&16&11&16&15\\
\bottomrule
\end{tabular}
\label{tab:euroroad_AEA_tvd}
\end{table}
%%%%%%%%%%%%%%%%%%%%%%%%%%%%%%%%%%%%%%%%%%
At this point it is worth stressing the fact that when a practitioner opts for the approximate exchange, they will not have a--priori knowledge of the number of auxiliary iterations needed to sample from the true target distribution. This means that they have two options: the first is to perform an experiment similar to the one for this example, where the total variation distance is calculated with respect to a
conservative MCMC run. The second option would be to proceed by choosing a large number of auxiliary iterations, eg. 500,000, and run the approximate exchange algorithm. However it is demonstrated that the latter can take up to 1.75 hours to run (Table \ref{tab:PL_timings_euroroad}), which turns out to be considerably more expensive from a computational point of view than the algorithm we propose.

\subsubsection{Pseudolikelihood-based Posterior Inference}\label{section:euroroad_PLMH}
Although fast (Table \ref{tab:PL_timings_euroroad}), sampling from the non--calibrated pseudo-posterior via the Metropolis-Hastings leads to considerably underestimated posterior standard deviations (Table \ref{tab:PL_euroroad}).

\begin{figure}[H]
    \centering
		\includegraphics[width=15cm,height=8cm]{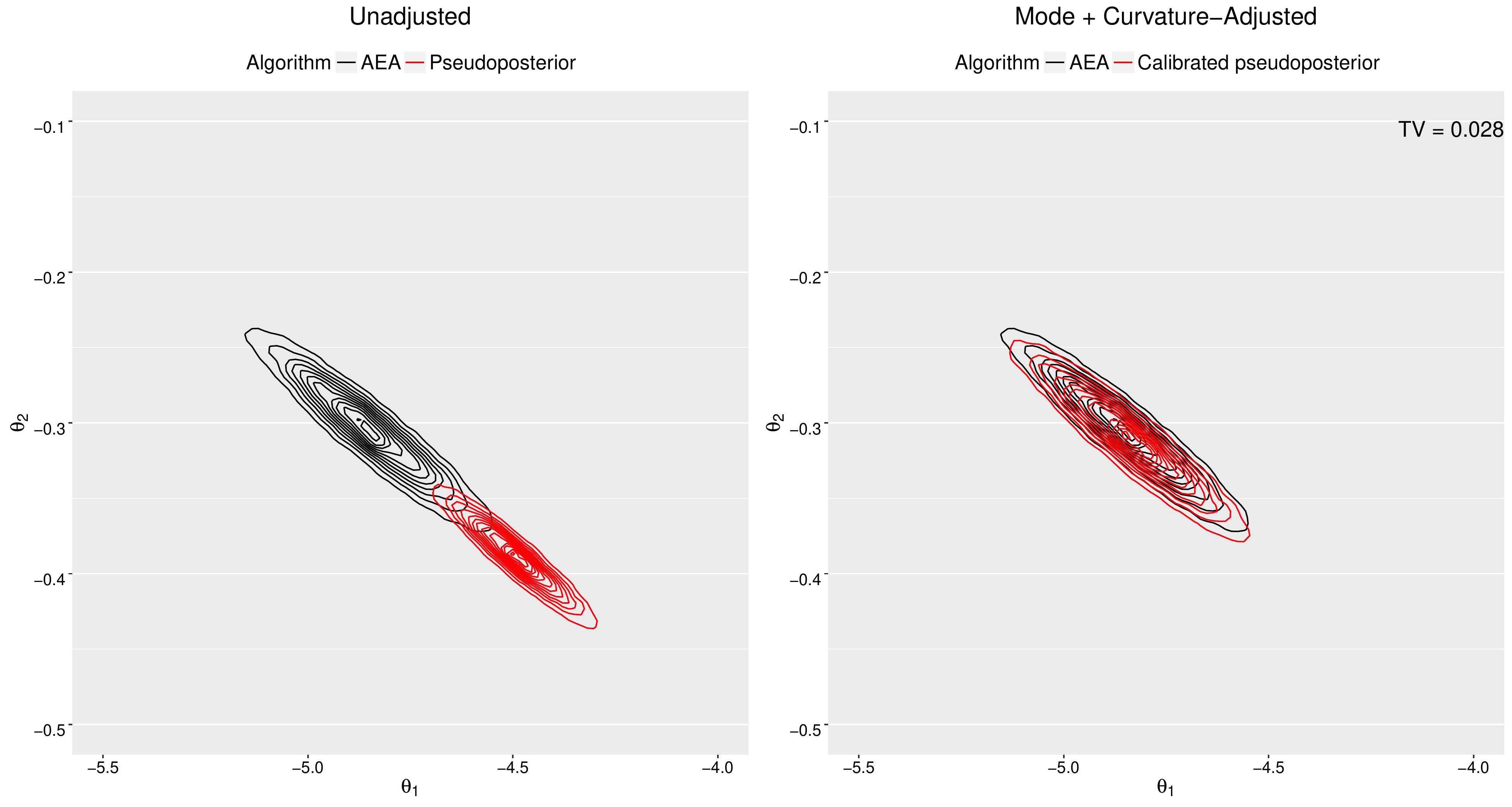}
		\caption{International E--road network: Phases of calibration of the misspecified posterior distribution using a pseudolikelihood approximation.}
\label{fig:PL_calibration_phases}
\end{figure}

The next step is to consider correcting the sample coming from the pseudo--posterior distribution. Monte Carlo estimation of the true posterior maximum can be performed in a quick manner. The Robbins–Monro stochastic algorithm converged after 50 iterations, where at each iteration 400 graphs were randomly drawn. Once the mode adjustment and the curvature adjustment were performed, a very good approximation of the true posterior with efficient correction of the posterior variance was obtained (Fig. \ref{fig:PL_calibration_phases}), while achieving a five--fold speedup relative to the approximate exchange with 10,000 auxiliary iterations.
%%%%%%%%%%%%%%%%%%%%%%%%%%%%%%%%%%%%%%%%%%
\begin{table}[H]
\vspace{-1.1em}
\caption{International E--road network: Pseudo--posterior parameter estimates (mean and standard deviation) at each phase of the calibration procedure.
The total variation distance was calculated between each joint posterior distribution and the joint posterior that was generated by the approximate exchange with 500,000 auxiliary iterations.}
\centering
\begin{tabular}{lrrr}
\toprule
& $\theta_1$ (Edges) & $\theta_2$ (2-stars) & TV  \\
\hline
Pseudo--posterior          &-4.496 (0.089) &-0.388 (0.021) &0.753\\
Mode + Curvature calibrated&-4.840 (0.127) &-0.311 (0.029) &0.028\\
\hline
 AEA ($10^4$ aux. iters)   &-4.846 (0.133) &-0.305 (0.030) &0.029\\
\bottomrule
\end{tabular}
\label{tab:PL_euroroad}
\end{table}
%%%%%%%%%%%%%%%%%%%%%%%%%%%%%%%%%%%%%%%%%%
The results of Table \ref{tab:euroroad_AEA_tvd} indicate that in the best case scenario, the average total variation distance will be 0.03 with respect to the long MCMC run. If one decides to use less iterations (eg. $5\times10^3$), this will result in a worse approximation with a higher total variation distance. On the other hand, after calibrating the
pseudo--posterior sample, the resulting total variation distance equals 0.028, showing a very good approximation to the true posterior. Any (slight) differences between the two distributions arise because of the Monte Carlo  error from the Robbins--Monro approximation. Hence, the overall calibration approach of the pseudo--posterior distribution provides a good trade--off between computational time and efficiency.

For an MCMC run of length T with lag k autocorrelation $\rho_{\text{k}}$ we measure the efficiency of the sampler by using the Effective Sample Size, $\text{ESS}=\text{T}/(1 + 2\sum\limits_{\text{k}=1}^{\infinity}\rho_{\text{k}})$ \citep{liu}. This Effective Sample Size (the larger the better) gives an estimate of the equivalent number of independent iterations that the chain represents. The algorithms considered in this paper have different running times; a fair comparison can be made by standardizing the ESS by the CPU run time, defining the efficiency ratio
%%%%%%%%%%%%%%%%%%%%%%%%%%%%%%%%%%%%%%%%%%
\[
\text{ER(Sampler)}=\frac{\text{Min ESS(Sampler)}}{\text{CPU(Sampler)}}
\]
%%%%%%%%%%%%%%%%%%%%%%%%%%%%%%%%%%%%%%%%%%
as a performance metric. To examine the performance of each algorithm relative to the ground truth approximate exchange sampler with 500,000 auxiliary iterations, we define the relative efficiency by
%%%%%%%%%%%%%%%%%%%%%%%%%%%%%%%%%%%%%%%%%%
\[
\text{RE}=\frac{\text{ER(Sampler)}}{\text{ER(AEA\textsubscript{500K})}}.
\]
%%%%%%%%%%%%%%%%%%%%%%%%%%%%%%%%%%%%%%%%%%
%%%%%%%%%%%%%%%%%%%%%%%%%%%%%%%%%%%%%%%%%%
The pseudolikelihood--based MH sampler can achieve a higher effective sample size compared to the approximate exchange (Table \ref{tab:PL_timings_euroroad}). Combined with the small overall computational cost of the calibration procedure, there is a huge gain in efficiency with respect to the ground truth approximate exchange algorithm, as well as the sampler with the 10,000 auxiliary iterations which were found to be practical for this example.
%%%%%%%%%%%%%%%%%%%%%%%%%%%%%%%%%%%%%%%%%%
\begin{table}[H]
\caption{International E--road network: CPU time (in seconds) of the different steps of the inference algorithms, minimum ESS, Efficiency Ratio and relative efficiency ratio. The total CPU time for the calibration procedure represents the sum of the CPU times of each of the individual stages.}
\centering
\begin{tabular}{lrrrrr}
\toprule
Calibration phase                & CPU (s)& ESS & ER & Relative ER\\
\hline
Pseudo--posterior                &14.36  &3638 &253.34 &816.23\\
Robbins--Monro (50 iters)        &11.86  &--    &--      &--\\
Mode + Curvature calibration     &8.87   &--    &--      &--\\
Total                            &35.09  &3638 &103.67 &333.41\\
\hline
 AEA ($10^4$ aux. iters)         &174.63 &1826 &10.45 &32.71\\
 AEA ($5\times10^5$ aux. iters)  &6297.96&1927 &0.31  &1    \\
\bottomrule
\end{tabular}
\label{tab:PL_timings_euroroad}
\end{table}
%%%%%%%%%%%%%%%%%%%%%%%%%%%%%%%%%%%%%%%%%%

\subsection{Faux Mesa High School Network}
\begin{figure}[H]
\centering
\includegraphics[width=9cm,height=9cm]{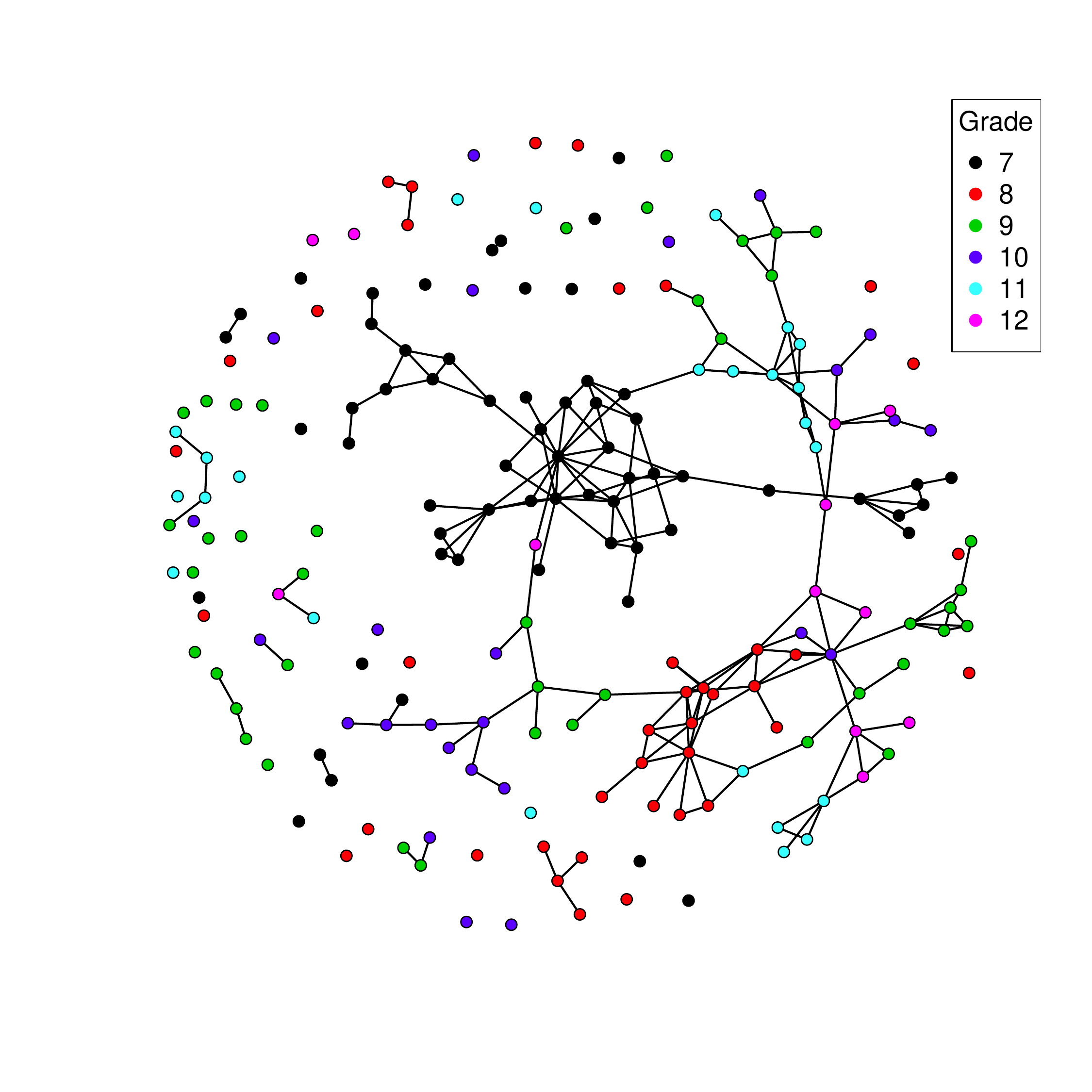}
\vspace{-2.0em}
\caption{Faux Mesa High School friendship network graph.}
\label{fig:faux_mesa_graph}
\end{figure}

The undirected graph displayed in Fig. \ref{fig:faux_mesa_graph} represents friendship relations in a school community of 205 students \citep{handcock:statnet}. The Faux Mesa High School network is a well known network in social science and consists of 203 undirected edges (mutual friendships). Similarly to \cite{caimo4}, we are interested in the vertex attributes \textit{x} for the "grade" (values 7 through 12 indicating each student's grade in school) of each student.
%%%%%%%%%%%%%%%%%%%%%%%%%%%%%%%%%%%%%%%%%%
\vspace{-1.1em}
%%%%%%%%%%%%%%%%%%%%%%%%%%%%%%%%%%%%%%%%%%
The main focus lies in the factor attribute effects (which give information about the tendency of a node with a specific attribute to form an edge in the network) and on the transitivity effect expressed by the GWESP statistics. We consider a model defined by the following 8 network statistics:
\begin{center}
\begin{tabular}{lr}
$s_{1}(y)  = \sum_{i<j}y_{ij}$&
$s_{5}(y,x)= \sum_{i<j}y_{ij} \{1_{(grade_{i} = 10)} + 1_{(grade_{j} = 10)}\}$\\[.1cm]
$s_{2}(y,x)= \sum_{i<j}y_{ij} \{1_{(grade_{i} = 7)}  + 1_{(grade_{j} = 7)}\}$&
$s_{6}(y,x)= \sum_{i<j}y_{ij} \{1_{(grade_{i} = 11)} + 1_{(grade_{j} = 11)}\}$\\[.1cm]
$s_{3}(y,x)= \sum_{i<j}y_{ij} \{1_{(grade_{i} = 8)}  + 1_{(grade_{j} = 8)}\}$&
$s_{7}(y,x)= \sum_{i<j}y_{ij} \{1_{(grade_{i} = 12)} + 1_{(grade_{j} = 12)}\}$\\[.1cm]
$s_{4}(y,x)= \sum_{i<j}y_{ij} \{1_{(grade_{i} = 9)}  + 1_{(grade_{j} = 9)}\}$&
$s_{8}(y)  = v(y,\phi_v)$,\\[.1cm]
\end{tabular}
\end{center}
where $1_{(\cdot)}$ %$1_{(\cdot)}$ 
is the indicator function. GWESP, the geometrically weighted edgewise shared partners, are given by the formula:
%%%%%%%%%%%%%%%%%%%%%%%%%%%%%%%%%%%%%%%%%%
\begin{align*}
v(y,\phi_v) = e^{\phi_v} \sum_{i=1}^{n-2}\left\{ 1-\left(1 - e^{-\phi_v} \right)^{i}\right\}EP_i(y),
\end{align*}
%%%%%%%%%%%%%%%%%%%%%%%%%%%%%%%%%%%%%%%%%%
\noindent where $EP_i(y)$ are the edgewise shared partners. To get a model which is a non--curved ERG model, we set $\phi_{v} = 1$. The pseudolikelihood--based Metropolis--Hastings sampler was tuned in the same way as described in Section \ref{section:euroroad} to obtain an overall acceptance rate of 24.3\%. Accordingly, each of the approximate exchange samplers
for the model of this example had an overall acceptance rate 23-30\%.

The experiment for this example were nodal attributes are accounted for consists of running the approximate exchange algorithm long enough (500,000 auxiliary iterations) to draw from the target distribution and then compare the posterior summary statistics with the corresponding quantities from the pseudo--posterior distribution (pre-- and post--calibration). We additionally compare the samplers in terms of the effective sample size and their respective efficiencies, as described below.
%%%%%%%%%%%%%%%%%%%%%%%%%%%%%%%%%%%%%%%%%%
\begin{table}[H]
\caption{Faux Mesa High School network: Posterior parameter estimates (mean and standard deviation) obtained by the approximate exchange sampler under an increasing number of auxiliary iterations.}
\centering
\begin{tabular}{lrrrr}
\toprule
 Auxiliary Iterations      & $\theta_1$ & $\theta_2$ & $\theta_3$ &$\theta_4$  \\
\hline
50             & -6.641 (0.757) & 2.613 (1.477) & 2.122 (2.411) & 2.495 (2.260) \\
100            & -6.561 (0.500) & 2.365 (0.863) & 2.076 (1.424) & 2.430 (1.347) \\
500            & -6.505 (0.252) & 2.251 (0.401) & 2.026 (0.563) & 2.308 (0.527) \\
$1\times10^3$  & -6.471 (0.204) & 2.156 (0.298) & 1.971 (0.447) & 2.204 (0.400) \\
$5\times10^3$  & -6.267 (0.166) & 1.875 (0.223) & 1.902 (0.288) & 1.944 (0.298) \\
$1\times10^4$  & -6.209 (0.162) & 1.855 (0.203) & 1.951 (0.247) & 1.921 (0.267) \\
$2\times10^4$  & -6.192 (0.166) & 1.897 (0.212) & 2.057 (0.239) & 2.000 (0.243) \\
$4\times10^4$  & -6.166 (0.171) & 1.926 (0.212) & 2.117 (0.233) & 2.045 (0.243) \\
$1\times10^5$  & -6.149 (0.196) & 2.012 (0.221) & 2.195 (0.239) & 2.058 (0.277) \\
$5\times10^5$  & -6.103 (0.177) & 2.052 (0.202) & 2.225 (0.221) & 2.051 (0.259) \\

\hline
\hline
\noalign{\vskip 0.02in}
Auxiliary Iterations       & $\theta_5$ & $\theta_6$ & $\theta_7$ &$\theta_8$  \\
\hline
50            & 3.886 (3.479) & 2.827 (3.526) & 3.497 (4.960) & 2.265 (0.989) \\
100           & 3.731 (2.115) & 2.713 (2.446) & 4.793 (3.884) & 1.572 (0.539) \\
500           & 3.015 (0.805) & 2.414 (0.897) & 4.258 (1.551) & 1.403 (0.214) \\
$1\times10^3$ & 2.913 (0.601) & 2.406 (0.637) & 4.082 (1.133) & 1.367 (0.158) \\
$5\times10^3$ & 2.544 (0.432) & 2.271 (0.409) & 3.625 (0.625) & 1.221 (0.106) \\
$1\times10^4$ & 2.371 (0.397) & 2.303 (0.337) & 3.479 (0.524) & 1.152 (0.092) \\
$2\times10^4$ & 2.285 (0.379) & 2.413 (0.284) & 3.295 (0.485) & 1.074 (0.078) \\
$4\times10^4$ & 2.218 (0.388) & 2.446 (0.281) & 3.119 (0.421) & 1.024 (0.073) \\
$1\times10^5$ & 2.210 (0.385) & 2.500 (0.274) & 2.979 (0.406) & 0.951 (0.065) \\
$5\times10^5$ & 2.213 (0.353) & 2.506 (0.251) & 2.839 (0.373) & 0.885 (0.059) \\
\bottomrule
\end{tabular}
\label{tab:faux_AEA_summary_convergence}
\end{table}
%%%%%%%%%%%%%%%%%%%%%%%%%%%%%%%%%%%%%%%%%%
The results of Table \ref{tab:faux_mesa} suggest that the network is very sparse (negative parameter $\theta_1$). Homophily is expressed by the positive posterior parameter means for the main effect of student grade and the positive GWESP parameter ($\theta_8$) expresses the transitivity effect. There is a clear tendency for the posterior standard deviations from the approximate exchange algorithm to decrease with an increasing number of auxiliary iterations (Table \ref{tab:faux_AEA_summary_convergence}). Overall, the posterior summary statistic values appear to stabilize when a few hundreds thousand auxiliary iterations are used.

The calibration procedure corrects the posterior standard deviations to a great extent (Table \ref{tab:faux_mesa}) and the respective posterior means are close to those obtained by the approximate exchange algorithm. A high--dimensional model such as the one examined in this example leads to increased computational time for the approximate exchange algorithm, if thousands of iterations are to be used in the auxiliary chain.
%%%%%%%%%%%%%%%%%%%%%%%%%%%%%%%%%%%%%%%%%%
\begin{table}[H]
\caption{Faux Mesa High School network: Posterior parameter estimates (mean and standard deviation).}
\centering
\begin{tabular}{lrrrr}
\toprule
& $\theta_1$ & $\theta_2$ & $\theta_3$ &$\theta_4$  \\
\hline
Pseudo--posterior              & -6.250 (0.163) & 1.805 (0.223) & 1.821 (0.281) & 2.090 (0.290) \\
Mode + Curvature calibrated    & -6.104 (0.150) & 2.051 (0.189) & 2.238 (0.219) & 2.061 (0.244) \\
AEA ($5\times10^5$ aux. iters) & -6.103 (0.177) & 2.052 (0.202) & 2.225 (0.221) & 2.051 (0.259) \\
\hline
\hline
\noalign{\vskip 0.02in}
 & $\theta_5$ & $\theta_6$ & $\theta_7$ &$\theta_8$  \\
\hline
Pseudo--posterior              & 2.353 (0.395) & 2.487 (0.331) & 2.827 (0.539) & 1.136 (0.053) \\
Mode + Curvature calibrated    & 2.208 (0.356) & 2.501 (0.218) & 2.859 (0.510) & 0.889 (0.082) \\
AEA ($5\times10^5$ aux. iters) & 2.213 (0.353) & 2.506 (0.251) & 2.839 (0.373) & 0.885 (0.059) \\
\bottomrule
\end{tabular}
\label{tab:faux_mesa}
\end{table}
%%%%%%%%%%%%%%%%%%%%%%%%%%%%%%%%%%%%%%%%%%
While we cannot provide a measure like the total variation distance to calculate the similarity between the joint posterior distributions under examination, we can compare the efficiency of the samplers. Table \ref{tab:PL_timings_school} illustrates that the analyst can benefit from a huge gain in terms of "information size" if the pseudolikelihood--based MH sampler is used. This, in combination with the short runtime of the calibration procedure, leads to increased efficiency relative to the long--run approximate exchange algorithm.
%%%%%%%%%%%%%%%%%%%%%%%%%%%%%%%%%%%%%%%%%%
\begin{table}[H]
\caption{Faux Mesa High School network: CPU time (in seconds) of the calibration procedure of the uncalibrated pseudo--posterior sample, minimum ESS and efficiency ratio. The total CPU time for the calibration procedure represents the sum of the CPU times of each of the individual stages.}
\centering
\begin{tabular}{lrrrr}
\toprule
Calibration stage            & CPU (s) & ESS & ER& Relative ER\\
\hline
Pseudo--posterior              &11.71&1554&132.78&44,259\\
Robbins--Monro  (50 iters)     &12.73&--  &--    &--     \\
Mode + Curvature calibration   &4.42 &--  &--    &--     \\
Total                          &28.86&1554&53.85 &17,949\\
\hline
 AEA ($1\times10^5$ aux. iters)&4,607.82 &130&0.028&8.33 \\
 AEA ($5\times10^5$ aux. iters)&40,212.43&124&0.003&1    \\
\bottomrule
\end{tabular}
\label{tab:PL_timings_school}
\end{table}
%%%%%%%%%%%%%%%%%%%%%%%%%%%%%%%%%%%%%%%%%%

%====================================================================================
%====================================================================================
\section{Discussion}\label{section:discussion}
In this paper we explored Bayesian inference of ERG models with tractable approximations to the true likelihood and we applied our methodology in real networks of increased complexity. The computational tractability of the pseudolikelihood function, which is algebraically identical to the likelihood for a logistic regression, has made it an attractive alternative to the full likelihood function. 
Bayesian logistic regression based on the pseudo-posterior distribution that results from replacing the true likelihood function with the pseudo-likelihood function can be carried relatively quickly. However the drawback of using such an approach is that it ignores strong dependencies in the data, since it involves the pseudolikelihood function and therefore can result in biased estimation. Our main contribution has been to demonstrate how to successfully correct a sample from the pseudo-posterior distribution so that it is approximately distributed from the target posterior distribution.

Our results showed that calibrating the pseudo-posterior distribution is a viable approach that outperforms the approximate exchange algorithm in terms of computational time and scales well to realistic-size problems, e.g  networks with around $1,000$ nodes. In comparison with the approximate exchange algorithm which is routinely used for the Bayesian analysis of ERGMs and is feasible for networks up to 1,000 nodes, our algorithm gave a dramatic reduction in CPU time and we anticipate it to scale well for even larger networks. 
With a two-step post-processing task we were able to perform a fast and efficient correction using the posterior mean and covariance. The immediate advantage of the proposed framework is evident when analyzing even larger networks and/ or more complex ERG models are fitted, where algorithms like the approximate exchange algorithm might struggle due to the size of the graph or the higher--dimensionality of the parameter vectors. In such cases where a comparison with the approximate exchange would be infeasible, we recommend applying our calibration algorithm.

Additional experiments were conducted using conditional composite likelihoods, which are higher-order generalizations of the pseudolikelihood function. In particular, we considered conditional composite likelihood components, or "blocks", consisting of three dyads each, conditioned on the remainder of the network. However the number of possible such blocks grows very quickly with $n$, and so one immediate difficulty was the issue of how to choose a subset of these in order to achieve a reasonable computational cost. 
Overall the results were not so promising. While the pseudolikelihood approach is a fast and viable method that uses all possible dyads, using the composite likelihood did not lead to a competitive sampler. Due to the complex nature of the random graph, difficulties arise in the systematic partition of observations into blocks and the selection of the subsets. Additionally, the overall computational cost can be prohibitive as composite likelihoods do not enjoy the fast point-wise estimation of pseudolikelihoods allowed by the change statistics reparameterization.

We hope that the arguments and findings presented here will serve as a set of guidelines for practitioners to draw meaningful inferences from relational data through the Bayesian paradigm, thus broadening the scope of network modeling beyond its current limits.

%====================================================================================
%====================================================================================
\section*{Acknowledgments}
The authors are grateful for the constructive feedback they received from the Editors and Reviewers at Social Networks. The Insight Centre for Data Analytics is supported by Science Foundation Ireland under Grant Number SFI/12/RC/2289. Nial Friel's research was also supported by a Science Foundation Ireland grant: 12/IP/1424.

%====================================================================================
%====================================================================================
%% References:
\bibliographystyle{apalike}
\bibliography{SON_1014}

\end{document}